\documentclass{aa}
\usepackage{epsf}

\begin{document}

\def\pFn{p_{\raise-0.3ex\hbox{{\scriptsize F$\!$\raise-0.03ex\hbox{\rm n}}}}
}  
\def\pFp{p_{\raise-0.3ex\hbox{{\scriptsize F$\!$\raise-0.03ex\hbox{\rm p}}}}
}  
\def\pFe{p_{\raise-0.3ex\hbox{{\scriptsize F$\!$\raise-0.03ex\hbox{\rm e}}}}
}  
\def\pFmu{p_{\raise-0.3ex\hbox{{\scriptsize F$\!$\raise-0.03ex\hbox{\rm
$\mu$}}}} }  
\def\m@th{\mathsurround=0pt }
\def\eqalign#1{\null\,\vcenter{\openup1\jot \m@th
   \ialign{\strut$\displaystyle{##}$&$\displaystyle{{}##}$\hfil
   \crcr#1\crcr}}\,}
\newcommand{\vp}{\mbox{\boldmath $p$}}         
\newcommand{\vS}{\mbox{\boldmath $S$}}
\newcommand{\vP}{\mbox{\boldmath $P$}}

\newcommand{\vk}{\mbox{\boldmath $k$}}         
\newcommand{\xixi}{\mbox{\boldmath $\xi$}}         
\newcommand{\vq}{\mbox{\boldmath $q$}}         
\newcommand{\vr}{\mbox{\boldmath $r$}}         

\newcommand{\om}{\mbox{$\omega$}}              
\newcommand{\Om}{\mbox{$\Omega$}}              
\newcommand{\Th}{\mbox{$\Theta$}}              
\newcommand{\ph}{\mbox{$\varphi$}}             
\newcommand{\del}{\mbox{$\delta$}}             
\newcommand{\Del}{\mbox{$\Delta$}}             
\newcommand{\lam}{\mbox{$\lambda$}}            
\newcommand{\Lam}{\mbox{$\Lambda$}}            
\newcommand{\ep}{\mbox{$\varepsilon$}}         
\newcommand{\ka}{\mbox{$\kappa$}}              
\newcommand{\dd}{\mbox{d}}                     
\newcommand{\vect}[1]{\bf #1}                
\newcommand{\vtr}[1]{\mbox{\boldmath $#1$}}  
\newcommand{\vF}{\mbox{$v_{\mbox{\raisebox{-0.3ex}{\scriptsize F}}}$}}  
\newcommand{\pF}{\mbox{$p_{\mbox{\raisebox{-0.3ex}{\scriptsize F}}}$}}  
\newcommand{\kF}{\mbox{$k_{\rm F}$}}           
\newcommand{\kTF}{\mbox{$k_{\rm TF}$}}         
\newcommand{\kB}{\mbox{$k_{\rm B}$}}           
\newcommand{\tn}{\mbox{$T_{{\rm c}n}$}}        
\newcommand{\tp}{\mbox{$T_{{\rm c}p}$}}        
\newcommand{\te}{\mbox{$T_{eff}$}}             
\newcommand{\ex}{\mbox{\rm e}}                 
\newcommand{\rate}{\mbox{${\rm erg~cm^{-3}~s^{-1}}$}}
\newcommand{\mur}{\raisebox{0.2ex}{\mbox{\scriptsize (Œ)}}} 
\newcommand{\Mn}{\raisebox{0.2ex}{\mbox{\scriptsize (Œ{\it n\/})}}}        %
\newcommand{\Mp}{\raisebox{0.2ex}{\mbox{\scriptsize (Œ{\it p\/})}}}        %
\newcommand{\MN}{\raisebox{0.2ex}{\mbox{\scriptsize (Œ{\it N\/})}}}        %

\title{ \bf Direct Urca Process in a Neutron Star Mantle}

\author{ M.~E.\ Gusakov \inst{1},  
         D.~G.\ Yakovlev \inst{1}
\and	 
         P.~Haensel \inst{2}
\and  
         O.~Y.\ Gnedin \inst{3}	 
	   }
\institute{
         Ioffe Physical Technical Institute,
         Politekhnicheskaya 26, 194021 St.~Petersburg, Russia,
        {\it gusakov@astro.ioffe.ru; yak@astro.ioffe.ru } 
        \and
        Copernicus Astronomical Center,
        Bartycka 18, 00-716 Warsaw, Poland, {\it haensel@camk.edu.pl}
        \and
        Space Telescope Science Institute,
        3700 San Martin Drive, Baltimore, MD 21218, USA,
	{\it ognedin@stsci.edu}
}
\offprints{M.E.\ Gusakov}
\date{Received xx xxxxx 2003 / Accepted xx xxxxx 2003}
\abstract
{
We show that the direct Urca process of neutrino 
emission is allowed in two possible phases 
of nonspherical nuclei 
(inverse cylinders and inverse spheres)
in the mantle of a neutron star near the crust-core
interface. The process is open because neutrons and
protons move in a periodic potential created by inhomogeneous
nuclear structures. In this way the nucleons acquire large quasimomenta
needed to satisfy momentum-conservation in the neutrino reaction. 
The appropriate neutrino emissivity 
in a nonsuperfluid matter is about 2--3 orders of magnitude higher
than the emissivity of the modified Urca process in the 
stellar core. The process may noticeably accelerate the
cooling of low-mass neutron stars. 
}
\titlerunning{Direct Urca process in a neutron star mantle}
\authorrunning{M.~E.~Gusakov, D.~G.~Yakovlev, P.~Haensel, O.~Y.~Gnedin}
\maketitle

\section{Introduction}
\label{introduction}

It is well known (Lattimer et al. \cite{lpph91}) that  
direct Urca  process produces the most
powerful neutrino emission in 
the inner cores of neutron stars (NSs). 
The simplest direct Urca process
in a dense degenerate matter, composed of neutrons (n)
with an admixture of
protons and electrons (p and e), 
consists of two successive reactions (direct and inverse ones):
\begin{equation}
{\rm n \to p + e} + \bar{\nu}_{\rm e}, 
\quad {\rm p + e \to n +} \nu_{\rm e},
\label{durca}
\end{equation}
where $\nu_e$ and $\bar{\nu}_{\rm e}$ are the electron 
neutrino and antineutrino, respectively.
The process is allowed by momentum conservation
if $\pFn < \pFp + \pFe$, and forbidden otherwise. 
Here, $\pFn, \pFp$, and $\pFe$ are the Fermi-momenta
of n, p, and e. 
The neutrino momentum $p_{\nu} \sim k_{\rm B}T$ is much
smaller than these Fermi momenta, and it can be neglected in 
momentum conservation 
($T$ is the internal NS temperature, and $k_{\rm B}$ is the 
Boltzmann constant).
It turns out that the direct Urca process 
is allowed only at sufficiently high densities
(typically, a few times of $\rho_0$, where
$\rho_0 = 2.8 \times 10^{14}$~g~cm$^{-3}$ is
the nuclear matter density at saturation) 
for those model equations of state of dense matter
which have a large symmetry energy (rather high fractions 
of protons and electrons) to satisfy the inequality
$\pFn < \pFp + \pFe$. 

Thus, the direct Urca process
is forbidden in the cores of low-mass
NSs. The main neutrino emission
from these (nonsuperfluid) cores is produced by the modified Urca
process (nN$\to$peN$\bar{\nu}_e$, peN$\to$nN$\nu_e$,
where N=n or p is a nucleon-spectator 
required for momentum conservation).
The modified Urca process is
6--7 orders of magnitude 
weaker than the direct Urca process. This greatly reduces the neutrino
emission
of low-mass NSs in comparison with the emission of massive NSs,
where the direct Urca is open.

In this paper we analyze the  
possibility to open the direct Urca process in
the inner NS crust. The main idea is that a momentum excess
may be absorbed by a lattice of nonuniform nuclear 
structures in the crust. Specifically, we consider a model
of nonspherical nuclear structures which appear
in the density range from $\rho \approx 10^{14}$~g~cm$^{-3}$
to the crust-core interface ($\rho_{\rm cc} \approx \rho_0/2$)
if one employs some models of nucleon-nucleon interaction
(Ravenhall et al.\ \cite{rpw83},
Pethick \& Ravenhall \cite{pr95}). The theory predicts 
a sequence of phase transitions
with increasing $\rho$ within this density range:
from familiar spherical nuclei to cylindrical nuclear structures,
from cylinders to slab-like structures, from slabs to
inverted cylinders, then to inverted spheres and, finally
(at $\rho=\rho_{\rm cc}$), to a uniform nuclear matter
in the core. The shell of nonspherical nuclei in the crust,
sometimes called the {\it NS mantle},
is thin (not thicker than several hundred meters) but
contains a noticeable fraction of the crust mass.    
We will consider two last phases -- 
the inverted cylinders and the inverted spheres,
where free protons appear (in addition to free neutrons
in the inner crust),
but periodic nuclear structures are still not dissolved
into the uniform nuclear matter. The
periodic structures modulate motion of neutrons and
protons (inducing Bloch states)
and open direct Urca process in a neutron-star mantle.

\section{Periodic potential and nucleon wave functions}
\label{potential}

In order to calculate the neutrino emissivity 
of direct Urca process in the NS mantle
we need wave functions of neutrons and protons in a 
periodic nuclear potential. For understanding the main
features of the problem, we adopt a simplified
Thomas-Fermi approximation. The nuclear structures
will be described using the results of Oyamatsu (\cite{oya93}).
Specifically, we will employ his model I for 
the energy density functional.
The potential energy of neutrons or protons
($j=$ n or p) can be calculated as
\begin{eqnarray}
  V_{j}(r) &=& 
  \frac{\partial}{\partial n_j} 
  \bigg\{ ( \epsilon_0(n_{\rm n}, n_{\rm p}) \bigg.
\nonumber \\
 && -  \left. \frac{3}{5} (3 \pi^2)^{2/3} 
 \left( 
 \frac{\hbar^2}{2 m_{\rm n}} n_{\rm n}^{5/3} 
 + \frac{\hbar^2}{2 m_{\rm p}} n_{\rm p}^{5/3}
 \right) \right\},
\label{pot}
\end{eqnarray}
where $\epsilon_0(n_{\rm n}, n_{\rm p})$ is the energy density of
the uniform nuclear matter, $m_j$ is the nucleon mass,
and $n_j(r)$  is a local number density of nucleon species $j$,
which depends on distance $r$ from the nucleus center inside 
a Wigner-Seitz cell [Eq.\ (4.8) of Oyamatsu \cite{oya93}]. 
Equation (\ref{pot})  neglects small
gradient corrections. For simplicity, we restrict
ourselves to scalar (spin-independent) nuclear potentials.

Using the Schr\"odinger perturbation theory 
we can express
the Bloch wave function of a nucleon (n or p) in the  
periodical potential $V_j(r)$ as:
\begin{equation}
 \Psi_{\vec{p}} \!=\! {\chi_s \over \sqrt{\rm V}} 
  \left(  \!
  {\rm e}^{i \, \vec{p} \vec{r}}  \!
  + \! \sum_{\vec{q} \neq \vec{0}} 
  V_{j \vec{q}}  
  { {\rm e}^{i \, \vec{p}' \vec{r}} 
  \over E_{\vec{p}} \!-E_{\vec{p}'} } \right) 
  \! \equiv \! { \chi_s  \over \sqrt{\rm V}} \sum_{\vec{q}} \! C_{\vec{q}} 
  {\rm e}^{i \, \vec{p}' \vec{r}}, 
\label{psi}  
\end{equation}
where $\vec{q}$ is an inverse lattice vector, 
$\vec{p}' = \vec{p}+ \vec{q}$, V is the normalization volume,
$C_{\vec{0}} = 1$, and 
$C_{\vec{q}} = V_{ j \vec{q}}/(E_{\vec{p}}-E_{\vec{p}'})$ 
for $\vec{q} \neq \vec{0}$. Furthermore,
$\chi_s$ is a unit spinor ($\chi_s \chi_{s'} = \delta_{s s'}$),
$s= \pm 1$ is the sign of the nucleon spin 
projection onto the quantization axis,
$E_{\vec{p}} = \vec{p}^2/2 m_j^*$ 
is the unperturbed energy, $\vec{p}_j$ is the momentum,
$m_j^*$ is the effective mass at the Fermi surface, 
and $V_{j \vec{q}}$ is 
a Fourier-transform of the potential $V_j(r)$:
\begin{equation}
   V_{j \vec{q}} = \frac{1}{{\rm V}_{\rm cell}} 
   \, \, \int_{\rm cell} \,
   V_j(r) \, e^{-i \, \vec{q} \vec{r}} \, \dd \vec{r},
\label{Vq}
\end{equation}
${\rm V}_{\rm cell}$ being the elementary-cell volume. 
An exact calculation of 
$V_{ j \vec{q}}$ is difficult because   
the configuration of the elementary cell is complicated.
However, the calculation can be simplified due to
the short-ranged character of nuclear forces. 
Consequently, the nucleon potential
is almost constant near the cell boundaries, where it can
be replaced by the boundary value $V_{\infty}$.
Introducing a simplified Wigner-Seitz cell of equivalent volume 
as defined by Oyamatsu (\cite{oya93}), p.~434, we find: 
\begin{eqnarray}
    V_{j \vec{q}} 
     &=&  \frac{1}{{\rm V}_{\rm WS}} \, \, \int_{\rm cell} \,
    ( V_j(r) - V_{\infty} ) \, 
    {\rm e}^{-i \, \vec{q} \vec{r}} \, \dd \vec{r}
\nonumber \\
   &\approx& \frac{1}{{\rm V}_{\rm WS}} \, \, \int_{\rm WS} \,
   ( V_j(r) - V_{\infty} ) \, 
   {\rm e}^{-i \, \vec{q} \vec{r}} \, \dd \vec{r},
\label{Vq1}
\end{eqnarray}
where we have used the identity
\begin{equation}
\int_{\rm cell} \,
    {\rm e}^{-i \, \vec{q} \vec{r}} \, \dd \vec{r} = 0.
\nonumber \\
\end{equation}
Now in Eq.\ (\ref{Vq1}) 
we can integrate over the simplified Wigner-Seitz cell
(a cylinder or a sphere for  
the phases of inverted cylinders and inverted spheres,
respectively).

Unless the contrary is indicated,
we will use the units, where
$\hbar = c = k_{\rm B} ={\rm  V} = 1$. 
Notice, that the perturbation expansion (\ref{psi}) fails 
(becomes singular) at Bragg's diffraction points
(at $|\vec{p}| = |\vec{p} + \vec{q}|$) which signals
the special importance of the band structure near these points. 
However, we will see that nucleons with such
``resonant'' wave functions do not contribute to 
the neutrino process of study.

\section{General formalism}
\label{formalism}
%
The emissivity $Q$ of the direct Urca process in the NS mantle
is calculated in the same way as in the NS core
(Lattimer et al.\ \cite{lpph91}). 
Using the notations from the review article by
Yakovlev et al.\ (\cite{ykhg01}), we obtain:
\begin{equation}
       Q = 2 \, \int \, \frac{\dd \vec{p}_{\rm n}}{(2 \pi)^3} \, \,
      \dd W_{i \rightarrow f} \,  \epsilon_{\nu}  \, 
      f_{\rm n} (1- f_{\rm p}) \, (1- f_{\rm e}),
\label{Qd}
\end{equation}
where $f_j$ is the Fermi-Dirac distribution 
of particle species $j$(=n, p, e),
$\varepsilon_{\nu}$ is the neutrino energy,
$\dd W_{i \rightarrow f}$ is the differential 
probability of neutron decay
[calculated with the
wave functions (\ref{psi})].
The overall factor 2 doubles 
the emissivity of the neutron decay to account for the
contribution of the inverse reaction
(assuming $\beta$-equilibrium).
After standard simplifications 
(e.g., Yakovlev et al.\ \cite{ykhg01})
and integration over orientations of the neutrino momentum, we get:
\begin{eqnarray}
   \dd W_{i \rightarrow f} \, \frac{\dd \vec{p}_{\rm n}}{(2 \pi)^3} &=&
   \frac{(2 \pi)^4}{(2 \pi)^{12}} \, 
   \sum_{\vec{q}_{\rm n}, \vec{q}_{\rm p}} \,
   \delta(\epsilon_{\rm n}-\epsilon_{\rm p}-\epsilon_{\rm e}
   -\epsilon_{\rm \nu}) \, \,
\nonumber \\
&\times& 
   \delta(\vec{p}'_{\rm n}-\vec{p}'_{\rm p}-\vec{p}_{\rm e}) \, \,
\nonumber \\
&\times& |M_{fi}|^2  \,\, 4 \pi \, \epsilon_{\rm \nu}^2 \,\, 
   \dd \epsilon_{\rm \nu}
   \prod_{j=1}^{3} p_{{\rm F}_j} \, m_j^* \,\, \dd \epsilon_j \, 
   \dd \Omega_j, 
\label{dW}
\end{eqnarray}
where $\epsilon_j$ is the energy of particles species $j$,
$\dd \Omega_j$ is the solid angle element in the direction of $\vec{p}_j$,
$|\vec{p}_j| = p_{{\rm F}_j}$ (i.e., nonperturbed particle momenta 
$\vec{p}_j$ in the momentum-conserving delta function are placed at
Fermi surfaces), and
$m_{\rm e}^{*} = \mu_{\rm e}$
($\mu_{\rm e}$ being the electron chemical potential).
Finally, 
\begin{equation}
  |M_{fi}|^2 = 2 \, G_{\rm F}^2 \, \cos^2 \theta_{\rm C} \, 
  |C_{\vec{q}_{\rm n}}|^2
   \, |C_{\vec{q}_{\rm p}}|^2 \,
  (f_{\rm V}^2 + 3 g_{\rm A}^2) 
\label{Mfi}
\end{equation}
is the squared matrix element, summed over particle spins and averaged over
directions of neutrino momentum. Here,
$G_{\rm F}=1.436 \times 10^{-49}$ erg cm$^3$ 
is the Fermi weak interaction constant,
$\theta_{\rm C}$ is the Cabibbo angle, 
$\sin \theta_{\rm C}=0.231$,
$f_{\rm V}=1$ is the vector interaction constant, and 
$g_{\rm A}=1.26$ is the Gamow-Teller 
axial-vector interaction constant. 

The leading term in the sum  (\ref{dW})
over inverse lattice vectors corresponds to
$\vec{q}_{\rm n} = \vec{q}_{\rm p} = 0$. However, in this
term
$\vec{p}'_{\rm n}= \vec{p}_{\rm n}$, $\vec{p}'_{\rm p} = \vec{p}_{\rm p}$ 
(see Eq.\ (\ref{psi}) and the discussion afterwards),
and the neutrino emission is forbidden 
($\pFn \geq \pFp + \pFe$)
by momentum conservation. Thus, the main contribution into the 
emissivity $Q$ comes from smaller terms with
either $\vec{q}_{\rm n} = 0$ or $\vec{q}_{\rm p} = 0$. 
The terms with
$\vec{q}_{\rm n}\neq 0$ and $\vec{q}_{\rm p} \neq 0$ are
even smaller and can be neglected. The retained terms
are constructed in such a way that 
the momentum-conserving delta function excludes the ``dangerous''
Bragg diffraction points.
For instance, one can easily show that for
$\vec{q}_{\rm n} = 0$ the Bragg diffraction condition 
$|\vec{p}_{\rm p}| = |\vec{p}_{\rm p} +
\vec{q}_{\rm p}|$ is incompatible with momentum conservation
in Eq.\ (\ref{dW}).  
 
Inserting Eq.\ (\ref{dW}) into Eq.\ (\ref{Qd}) and 
integrating over particle
energies and propagation directions, we get:
\begin{eqnarray}
  Q(T,\rho)& = &Q_0(T,\rho) \, {\rm R}(\rho), 
\label{Qd1} \\
  Q_0(T,\rho) \!&=&\!\frac{457 \pi}{10080} \, 
  G_{\rm F}^2 \, \cos^2 \theta_{\rm C} \,
  (f_{\rm V}^2 + 3 g_{\rm A}^2) \, 
  m_{\rm n}^* m_{\rm p}^* m_{\rm e}^* \, T^6
\nonumber \\   
     & \approx& 4.0 \times 10^{27} \left( n_{\rm e} 
    \over n_0 \right)^{1/3} 
    { m_{\rm n}^* m_{\rm p}^* \over m_{\rm n} m_{\rm p}} \, T_9^6
    \;\;{\rm erg \over cm^3~s}, 
\label{Qd0} \\
  {\rm R}(\rho)\! &=& \!\!\!\sum_{j={\rm n, p}} \,\,\, 
  \sum_{\vec{q}}
\, \frac{(m_{j}^* \, V_{j\vec{q}})^2}{\alpha_j \,
  p_{{\rm F}_j}^4} \, \, 
\nonumber \\
  &&\!\!\!\!\times \!\biggl[ F(2 \alpha_j D^{\rm max}_j + \alpha_j^2) 
  - F(2 \alpha_j D^{\rm min}_j + \alpha_j^2) \biggr],
\label{R} 
\end{eqnarray}
where $n_{\rm e}$ is the electron
number density, $n_0=0.16$ fm$^{-3}$
is the nucleon number density in saturated nuclear matter, 
$T_9=T/10^9~{\rm K}$,
$\alpha_j = q/ p_{{\rm F}_j}$, 
$D^{\rm min}_j = {\rm max}(-1, \, D_{j-})$, 
$D^{\rm max}_j = {\rm min}(1, \, D_{j+})$,
\begin{eqnarray}
   F(x) &=&  \frac{1}{2} \, \ln
   \left| \frac{\sqrt{1+x}+1}{\sqrt{1+x}-1} \right|
   - \frac{\sqrt{1+x}}{x} \,,
\nonumber \\
   D_{{\rm n}\pm} &=& \frac{\,(\pFp \pm \pFe)^2 
   - \pFn^2 - q^2}{2 \pFn q}, 
\nonumber \\
   D_{{\rm p}\pm} &=& \frac{\,(\pFn \pm \pFe)^2 
   - \pFp^2 - q^2}{2 \pFp q}, 
\nonumber 
\end{eqnarray}
with $\pFn - \pFp - \pFe \leq q \leq \pFn + \pFp + \pFe$
[and $F(x)=0$ otherwise]. 
In Eq.\ (\ref{Qd1}) $Q_0$ is the direct Urca emissivity
in the uniform nuclear matter 
(disregarding the momentum-conservation constraint)
and R can be called its {\it reduction factor} which describes
the weakening of the process in the NS mantle. 
This factor depends on
$\rho$, but not on $T$. The emissivity $Q$
is $\propto T^6$, just as for the well-known direct Urca
process in the NS core.

For calculating R from Eq.\ (\ref{R}) 
we need the particle Fermi-momenta.
Because our analysis is simplified, we have used two models.
{\it First}, we have defined the Fermi-momenta by:
\begin{equation}
  \pFn = (3 \pi^2 n_{\rm n})^{1/3}, 
  \quad \pFp = \pFe = (3 \pi^2 n_{\rm p})^{1/3},
   \label{p1}
\end{equation}
where $n_{\rm n}$ and $n_{\rm p}$ are  
the nucleon number densities averaged over the Wigner-Seitz cell
[see Eq.\ (4.8) and Table 6 of Oyamatsu \cite{oya93}],
and $\pFp = \pFe$ due to electric neutrality.
{\it Second}, we have adopted electric neutrality 
and $\beta$-equilibrium,
and determined the Fermi-momenta from the equations:
\begin{equation}
  \pFe = \pFp,
  \quad \frac{\pFn^2}{2 m_{\rm n}} + V_{{\rm n} \vec{0}} =	
  \frac{\pFp^2}{2 m_{\rm p}} + V_{{\rm p} \vec{0}} + \pFe,
\label{p2}
\end{equation}
where $V_{j \vec{0}}$ is the central ($\vec{q}=\vec{0}$) Fourier harmonics 
of the nucleon potential $V_j(r)$.

The values of ${\rm R}(\rho)$ calculated from Eq.\ (\ref{R}) 
appear to be qualitatively the same for both models.
In our calculations we have set $m_j^*=m_j$,
but variations of $m_j^*$ within reasonable limits
do not qualitatively change
${\rm R}(\rho)$. The results based on the model (\ref{p2}) 
can be fitted by
\begin{eqnarray}
   {\rm R}(\rho) &\approx& {\rm R}_2 + ({\rm R}_1-{\rm R}_2) (1-x)^2 
   \quad  x<1,
\nonumber \\
    {\rm R}(\rho) &\approx& {\rm R}_2/x^7 \quad x \geq 1,
\label{Rfit}
\end{eqnarray}
where ${\rm R}_1 = 6 \times 10^{-5}$, ${\rm R}_2 = 10^{-5}$, 
$x=(n_{\rm b}-n_1)/(n_2-n_1)$,
$n_{\rm b}=\rho/m_{\rm n}$ is the baryon number density, while
$n_1$ and $n_2$ are the baryon number densities at the outer
and inner boundaries of the inverted-cylinder phase. Thus, the
density parameter $x$ varies from 0 to 1 in the layer of
inverted cylinders, and we have $x>1$ in the layer of inverted
spheres. In the Oyamatsu model, which we employ, 
$n_1=0.08274$ fm$^{-3}$,
$n_2=0.08537$ fm$^{-3}$, 
and the layer of inverted spheres extends to
$n_3 = 0.08605$ fm$^{-3}$ (to $x=4.868$). 

The number $N$
of inverted lattice vectors, which contribute into R, is large: 
$N \sim 200$ for the phase
of inverted cylinders, and $N$ is up to $\sim 2800$ for the phase
of inverted spheres.
The proton contribution ($\vec{q}_{\rm p} \neq \vec{0}$)
into R is approximately three times
larger than the neutron one ($\vec{q}_{\rm n} \neq \vec{0}$).

As seen from Eq.\ (\ref{Rfit}), $R \sim 10^{-5}$. 
Thus, the emissivity of the direct Urca process in the mantle
is about 5 orders of magnitude weaker than 
in the inner NS core.
Nevertheless, as will be shown in the next section, the direct Urca 
in the mantle can affect the cooling of NSs.

\begin{figure}[t]
\begin{center}
\epsfysize=80mm
\epsffile[80 215 560 680]{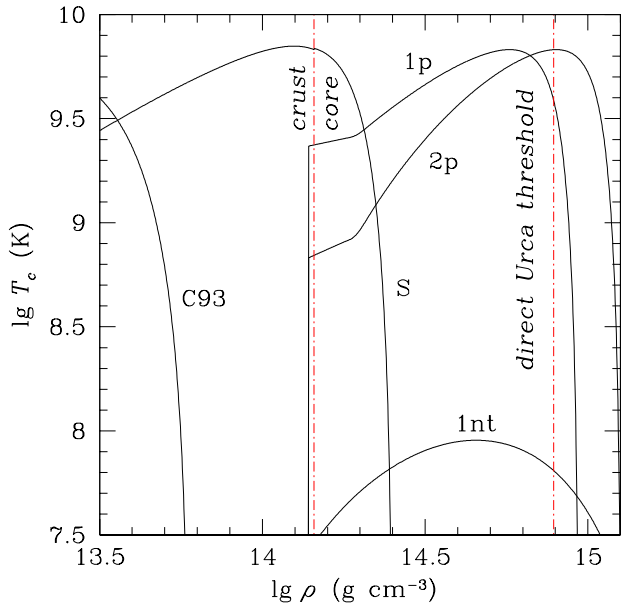}
\caption{
Critical temperature $T_{\rm c}$ versus $\rho$ for
various models of neutron and proton pairing in the NS core and crust.
Models C93 and S: 
singlet-state neutron pairing in the crust; 
models 1p and 2p: singlet-state proton pairing in the core; 
model 1nt: weak triplet-state neutron pairing in the core.
The crust-core interface and the threshold density 
of the direct Urca process in the core are shown
by vertical dot-and-dashed lines.
}
\end{center}
\label{gap}
\end{figure}

\section{The effect on the cooling of low-mass NSs}
\label{effect}
 
We will focus on sufficiently low-mass NSs,
with the forbidden direct Urca process in  
the inner cores (Sect.\ \ref{introduction}).
The cooling behaviour of these stars is described,
for instance, by Potekhin et al.\ (\cite{pycg03}).
The neutrino luminosity $L_\nu$ of low-mass stars
is not too high. Thus, the additional neutrino
emission from the mantle may be pronounced at the
neutrino cooling stage (when the stellar age  
$t \la 10^5$ yrs). 

\begin{figure}[t]
\begin{center}
\epsfysize=85mm
\epsffile[80 220 550 670]{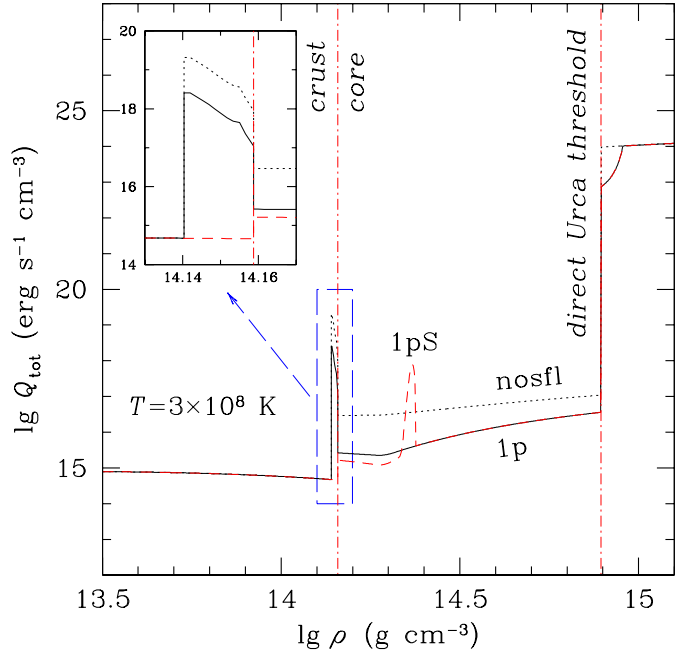}
\caption{
Total neutrino emissivity
versus density at $T = 3 \times 10^8$ K. 
The crust-core interface and the direct Urca threshold
in the core are marked by dot-and-dashed lines.
The peak near the core-crust interface 
shows the direct Urca process in the mantle
(presented also in the insert).
Dotted line: no nucleon superfluidity (nosfl);
solid line: model 1p of proton superfluidity; 
dashed line: model 
1p of proton superfluidity 
in the core and model S of neutron superfluidity 
in the crust (see the text).
The neutron pairing 1nt is too weak to appear 
and affect the neutrino emission at the given $T$.
}
\end{center}
\label{emis}
\end{figure}

An order-of-magnitude estimate gives
\begin{equation}
L_\nu \sim 4 \, \pi\, R_{\rm cc}^2 \, h  \,  Q\, +  
(4 \pi/ 3) \, R_{\rm cc}^3 \, Q_{\rm core}, 
\label{L}
\end{equation}
where $R_{\rm cc}$ is the core radius, $h$ is the mantle
width, and $Q_{\rm core}$ is the mean core neutrino emissivity.
Adopting $R_{\rm cc}=10$ km and $h=100$ m, we obtain that
the direct Urca in the mantle may affect the cooling if
$
Q \ga  30 \, Q_{\rm core}
$,
which is quite possible.
For instance, this inequality may hold for non-superfluid
low-mass NSs. The main neutrino emission 
in the cores of these NSs is produced by the
modified Urca process, which is 6--7 orders of magnitude weaker
than the direct Urca process in the cores of massive NSs,
while the direct Urca process in the mantle is five
orders of magnitude weaker than in the cores of massive NSs. 

Let us illustrate these statements by more elaborated
calculations. For this purpose we employ 
the equation of state of Prakash et al.\ (\cite{pal88})
in the stellar cores
(the version with the compression modulus of the saturated symmetric
nuclear matter $K=240$ MeV and with model I for the symmetry
energy). The NS models based on this equation of state
are described, e.g., by Yakovlev et al.\ (\cite{ykhg01}). 
The equation of state opens direct Urca process in the
NS core at $\rho \geq 7.85 \times 10^{14}$ g cm$^{-3}$
(which is possible
in NSs with $M>1.358$ $M_\odot$). The most massive stable
NS has the central density $\rho_{\rm c}=2.578 \times 10^{15}$
g cm$^{-3}$ and the mass $M=1.977$ $M_\odot$. 
A typical low-mass NS with $M=1.35\,M_\odot$ has $\rho_{\rm c}=7.79 \times
10^{14}$ g cm$^{-3}$ and $R=13.0$ km. 

We have simulated the NS cooling with our fully relativistic 
cooling code (Gnedin et al.\ \cite{gyp01}). It calculates
the cooling curves: the NS surface temperature $T_s^{\infty}$,
as detected by a distant observer, versus the NS age $t$. 
We have updated the code by incorporating the neutrino
emission from the direct Urca process in the NS mantle.
We have considered nonsuperfluid NSs and
NSs with neutron and proton superfluidity of the internal layers.
We have taken into account ({\it i})~a possible singlet-state pairing
of free neutrons in the crust and the outermost part
of the core, ({\it ii})~a triplet-state pairing of neutrons in the core,
and ({\it iii})~a singlet-state pairing of protons in the crust. 

Microscopic theories of nucleon superfluidity of dense matter
give very model dependent density profiles of superfluid
critical temperatures of nucleons, $T_{\rm c}(\rho)$
(e.g., Lombardo \& Schulze \cite{ls01}). Thus,
we have considered several superfluid models 
(Fig.\ 1) available in the literature:
one phenomenological model 1p of strong singlet-state
proton pairing, and one phenomenological model 1nt of
weak triplet-state neutron pairing in the NS core
(Kaminker et al.\ \cite{kyg02});
models S (Schulze et al.\ \cite{S}) and
C93 (Chen et al.\ \cite{C93}) of singlet-state
neutron pairing in the crust.
We have also proposed the additional phenomenological model 2p of
proton pairing in the core (not to be confused with model 2p
in Kaminker et al.\ \cite{kyg02}!).
This pairing is rather weak at the core-crust interface
but becomes much stronger at higher $\rho$.

The effects of superfluidity on the neutrino emission and
heat capacity of the matter have been included in the standard
way (Yakovlev et al.\ \cite{yls99}). 
It is well known that the nucleon superfluidity
reduces the traditional neutrino
mechanisms but opens the neutrino emission
due to Cooper pairing of nucleons.
We assume that
free protons and free neutrons in 
the mantle have the same 
$T_{\rm c}(\rho)$ as protons in the core and free neutrons
in the ordinary crust of spherical nuclei. 
We have adopted the same form of the superfluid
reduction factor of the direct Urca 
process in the mantle as in the inner core.

%
\begin{figure}[t]
\begin{center}   
\epsfysize=85mm 
\epsffile[18 144 577 680]{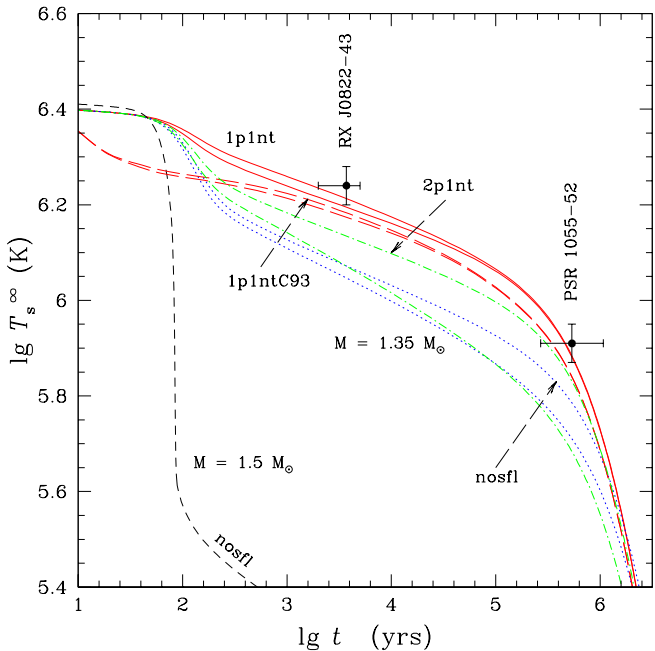}
\caption{
Theoretical cooling curves compared with observations
of two NSs. All the lines but the short-dashed 
one are for $1.35\,M_\odot$ NS
(direct Urca in the core forbidden).
Any pair of lines of the same type refers to one
superfluidity model with the direct Urca process in the mantle on
(lower curve) or off (upper curve).
Dotted curves: nonsuperfluid NSs (nosfl);
solid curves: 1p and 1nt superfluidities in the core;
long dashes: 1p and 1nt superfluidity in the 
core and C93 superfluidity of neutrons in the crust; 
dot-and-dash lines: 2p and 1nt superfluidities in the core.
The short dashed line is for a nonsuperfluid $1.5\,M_\odot$ NS
with the open direct Urca process in the core.
}
\end{center}
\label{cool}   
\end{figure}

Figure 2 shows the density profile of the total neutrino
emissivity at $T=3 \times 10^8$ K for
nonsuperfluid and superfluid NSs, while
Fig.\ 3 presents some cooling curves.
All the curves, but one short-dashed curve, are calculated
for a $1.35\,M_\odot$ NS with the forbidden direct Urca process in the core.
This is a typical example of a low-mass NS.
Any pair of lines of the same type corresponds to the same
superfluidity model. 
Every upper line of the pair is calculated
neglecting the direct Urca process in the mantle while
its lower counterpart is calculated including this process.
The short dashed curve is an example 
of the cooling of a more massive,
$1.5 \, M_\odot$, nonsuperfluid NS. Its central density is
$\rho_{\rm c} = 9.0 \times 10^{14}$ g cm$^{-3}$;
the direct Urca process is open in the inner stellar
core, producing a strong neutrino emission and a rapid
cooling of the star. In this case the direct Urca process in the mantle
is entirely negligible: it cannot compete 
with its senior partner in the core.  

Along with the theoretical curves, in Fig.\ 3
we present the observational limits on the effective
surface temperatures and ages of two NSs,
RX J0822--4300 and PSR B1055--52. Among all middle-aged isolated
NSs, whose thermal emission has been observed
and the effective surface temperature measured
(the data are given, e.g., in Yakovlev et al.\ \cite{cospar}), 
these are the NSs hottest for their ages.  
They can be interpreted as low-mass NSs
(see, e.g., Yakovlev et al.\ \cite{cospar}
and references therein). 

RX J0822--4300 is a radio silent NS, a 
compact central object in Puppis A.
Its effective temperature is taken
from Zavlin et al.\ (\cite{ztp99}). 
Recently Pavlov (\cite{pavlov03}) has kindly provided us with
the updated value of the effective temperature of PSR B1055--52,
$T_\mathrm{s}^\infty \approx 7 \times 10^{5}$ K.
However, he has not indicated errorbars. We introduce (Fig.\ 3),
somewhat arbitrarily, 10\% uncertainties of $T_\mathrm{s}^\infty$. 
The ages of RX J0822--4300 and PSR B1055--52
are taken as described by Yakovlev et al.\ (\cite{cospar}).

The dotted line in Fig.\ 2 shows the neutrino emissivity
in a nonsuperfluid NS.
The peak before the crust-core interface is produced
by the direct Urca process 
in the mantle (additionally presented in the insert).
The emissivity at the peak maximum is
about three orders of magnitude larger than in the NS core. 
The appropriate cooling curves are shown by the dotted
lines in Fig.\ 3. The direct Urca process in the mantle
noticeably accelerates the cooling at $t \sim 10^5$ yrs.

Assuming the strong 1p proton pairing and
the weak 1nt triplet-state neutron pairing in the core (and the mantle)
but no singlet-state neutron pairing in the crust, we
obtain the emissivity profile plotted by the solid line in Fig.\ 2.
The proton superfluidity fully suppresses the modified
Urca process in the outer core, before the direct Urca
threshold, and partly suppresses the direct Urca process
in the inner core (of massive NSs) beyond the threshold. 
The neutron pairing in the core is so weak that
it does not appear at $T = 3 \times 10^8$ K in Fig.\ 2. 
The neutrino emission from the core of a low-mass NS
becomes very slow, being mainly determined by the neutron-neutron
bremsstrahlung process. This increases the surface temperatures
of middle-aged NSs (the solid curves in Fig.\ 3), 
and enables one to interpret
the observations of RX J0822--4300 and PSR B1055--52
(e.g., Kaminker et al.\ \cite{kyg02}, 
Yakovlev et al.\ \cite{cospar}). In this case the
proton superfluidity partly suppresses the direct Urca process
in the mantle and reduces the difference of the cooling
curves calculated with and without this process.
Notice that the combination of the superfluidities
1p and 1nt is sufficient to explain the
data on all isolated middle-aged NSs [whose surface
temperatures have been measured (estimated) from
the observations of their thermal radiation] by 
theoretical cooling curves of NSs with different masses
(e.g., Kaminker et al.\ \cite{kyg02}). We see that the
inclusion of the direct Urca process in the mantle does not
violate this interpretation.

If we additionally switch on the singlet-state
superfluidity C93 in the crust, we will obtain
the long-dashed cooling curves in Fig.\ 3, which go
slightly lower than the solid curves.
This additional acceleration of the cooling is produced
by the neutrino emission due to Cooper-pairing of neutrons
in the crust. The effect is weak because
the superfluidity C93 dies out long before the
crust-core interface and produces the weak Cooper-pairing emission 
(Potekhin et al.\ \cite{pycg03}).
The relative importance of the direct Urca process
in the mantle is even lower than in the absence
of the superfluidity C93.

If we add the singlet-state neutron superfluidity S instead of C93,
the situation would be different in two respects.
First, the superfluidity S completely switches off  
the direct Urca process in the mantle (see the dashed
curve in Fig.\ 2; its peak
in the core is produced by the neutrino emission 
due to the pairing S). 
Second, the superfluidity S occupies much
larger fraction of the NS volume than the superfluidity C93, 
intensifying the neutrino emission
due to the singlet-state pairing of neutrons.
Now this emission noticeably lowers the cooling curve,
complicating the interpretation of RX J0822--4300
and PSR B1055--52. We do not present the cooling curve
for that unlikely case in Fig.\ 3; it is
the same as  given by Potekhin et al.\ (\cite{pycg03}).
 
Finally, let us employ the proton pairing 2p and
the neutron pairing 1nt in the core (and neglect the singlet-state
neutron pairing in the crust). 
The proton superfluidity 2p suppresses the neutrino emission
in the main fraction of the core but not in the mantle.
The effect of the direct Urca process in the mantle on the cooling of
NSs of the age $t \sim (10^4-10^5)$ yrs becomes most
pronounced (the dot-and-dash curves in Fig.\ 3).
With this process on (the lower dot-and-dash curve)
the low-mass star will cool too fast, strongly complicating
the interpretation of the observations of RX J0822--4300
and PSR B1055--52.

Let us emphasize that the existence of the NS mantle
(the layer of nonspherical atomic nuclei) is still a
hypothesis. The theory predicts this layer only within
some models of nucleon-nucleon interaction.
Since the lower dot-and-dash curve in Fig.\ 3 strongly contradicts the
observations, the underlying physical scenario
becomes doubtful. This would imply, for instance, that the
2p proton superfluidity model is inadequate, or
the neutrons in the mantle are strongly superfluid
(switching off the direct Urca process),
or the mantle does not exist at all.

Our analysis of cooling low-mass NSs is illustrative
and incomplete.
The cooling of these NSs is actually
affected by ({\it i}) NS superfluidity, 
({\it ii}) NS surface magnetic fields, and 
({\it iii}) possible surface layer
of light (accreted) elements (as discussed in detail
by Potekhin et al.\ \cite{pycg03}). Our calculations
indicate the existence of the fourth regulator, 
({\it iv}) the presence of the NS mantle and the associated
direct Urca process. As clear from the results of
Potekhin et al.\ (\cite{pycg03}) and our present results, 
{\it all four regulators are
of comparable strength} and should be analyzed together.
This many-parametric analysis is beyond 
the scope of the present article.

\section{Conclusions}
\label{conclusions}

We have calculated the neutrino emissivity of the new
neutrino mechanism -- the direct Urca process in a 
neutron-star mantle, a thin layer of nonspherical nuclei
(Ravenhall et al.\ \cite{rpw83}, Pethick \& Ravenhall \cite{pr95})
adjacent to the stellar core. The mantle is predicted only by 
some models of nucleon-nucleon interaction (and is
not predicted by other models). Thus, the existence
of the mantle is hypothetical. If exists, it cannot
noticeably affect the equation of state, and
the hydrostatic NS structure (particularly, NS masses
and radii). We expect that the strongest manifestation of the
mantle consists in opening direct Urca process. 
It can operate
in the two last phases of nonspherical nuclei 
(inverted cylinders and inverted spheres),
where the continuum proton spectrum is formed (e.g.,
Oyamatsu \cite{oya93}). The emissivity of the
new process in a nonsuperfluid mantle
appears to be 2--3 orders of magnitude higher
than the neutrino emissivity in the nonsuperfluid outer
NS core. 

We have performed illustrative calculations of
NS cooling which show that the new process
can noticeably affect the cooling of low-mass NSs. Its effect
is most pronounced in NSs with strongly superfluid cores
(to reduce the neutrino emission from the cores)
and nonsuperfluid mantles (to fully open direct Urca
process there). Thus, direct Urca process
in the mantle represents a new regulator of the cooling
of low-mass NSs.

Our calculation of the emissivity of the new process
is simplified (based on the Thomas-Fermi model with
a simplified form of scalar nucleon interaction, and
an approximate choice of nucleon Fermi momenta). 
The calculation can be improved but we expect that
the main results will be qualitatively the same.
One cannot exclude (Jones \cite{jones}) 
that direct Urca process
operates also in the crust of spherical atomic nuclei,
or in some selected layers of the crust, but its calculation
is difficult (requires exact wave functions
of nucleons). If operates, it could be a stronger
regulator of NS cooling than direct Urca process
in the mantle. 

It is important that delicate properties
of subnuclear matter can potentially be tested
by observations of cooling NSs. As clear from our
discussion, NSs hottest for their ages 
are the most suitable targets of such tests.

\begin{acknowledgements}
This work was supported in part by the
RFBR (grants 02-02-17668 and 03-07-90200),
the RLSS (grant 1115.2003.2),
KBN (grant 5 P03D 020 20), and by the INTAS (grant YSF 03-55-2397).
\end{acknowledgements}

\end{document}